\documentclass[prl,twocolumn,floatfix,groupedaddress,nofootinbib,showpacs,preprintnumbers,
amsmath,amssymb,amsfonts,widetable]{revtex4}
\usepackage{bm}
\usepackage{mathrsfs}
\usepackage{amssymb}
\usepackage{amsmath}
\usepackage{graphicx}
\usepackage{color}
\usepackage{array}

\begin{document}
\title{Quantifying Correlations Between Isovector Observables and the Density Dependence of Nuclear Symmetry Energy away from Saturation Density}
\author{F.~J. Fattoyev}\email{farrooh.fattoyev@tamuc.edu}
\author{W.~G. Newton}\email{william.newton@tamuc.edu}
\author{Bao-An Li}\email{bao-an.li@tamuc.edu}
\affiliation{Department of Physics and Astronomy, Texas A\&M
               University-Commerce, Commerce, TX 75429, USA}
\date{\today}
\begin{abstract}
\noindent According to the Hugenholtz-Van Hove theorem, the nuclear
symmetry energy $S(\rho)$ and its slope $L(\rho)$ at arbitrary
densities can be decomposed in terms of the density and momentum
dependence of the single-nucleon potentials in isospin-asymmetric
nuclear matter which are potentially accessible to experiment. We
quantify the correlations between several well-known isovector
observables and $L(\rho)$ to locate the density range in which each
isovector observable is most sensitive to the density dependence of
the $S(\rho)$. We then study the correlation coefficients between
those isovector observables and all the components of the $L(\rho)$.
The neutron skin thickness of $^{208}$Pb is found to be strongly
correlated with the $L(\rho)$ at a {\sl subsaturation} density of
$\rho = 0.59 \rho_0$ through the density dependence of the
first-order symmetry potential. Neutron star radii are found to be
strongly correlated with the $L(\rho)$ over a wide range of {\sl
supra-saturation} densities mainly through both the density and
momentum dependence of the first-order symmetry potential.  Finally,
we find that although the crust-core transition pressure has a
complex correlation with the $L(\rho)$, it is strongly correlated
with the momentum derivative of the first-order symmetry potential,
and the density dependence of the second-order symmetry potential.
\end{abstract}
\smallskip
\pacs{
21.65.Cd,       
21.65.Mn,       
21.65.Ef,       
26.60.Kp        
}
\maketitle

Improving knowledge of the density dependence of the nuclear
symmetry energy is an active endeavor due to its multifaceted impact
in many areas of nuclear physics and
astrophysics~\cite{Li:2014nse,Steiner:2004fi,Baran:2004ih,Lattimer:2006xb,Li:2008gp},
as well as in some issues regarding possible new physics beyond
standard
model~\cite{Horowitz:1999fk,Sil:2005tg,Krastev:2007en,Wen:2009av,Lin:2013rea}.
Despite intensive efforts aimed at constraining the density
dependence of the nuclear symmetry energy on both the experimental
and theoretical
fronts~\cite{Lattimer:2012xj,Lattimer:2012nd,Li:2013ola}, its
knowledge still remains largely uncertain even around nuclear
saturation density $\rho_0$~\cite{Fattoyev:2013yaa}. Traditionally,
the nuclear symmetry energy is expanded around $\rho_0$ as $S(\rho)
= J + L \chi + \frac{1}{2} K_{\rm sym} \chi^2 + \mathcal{O}(\chi^3)$
with $\chi \equiv (\rho-\rho_0)/3\rho_0$, and then individual
parameters of this expansion---particularly $J$ and $L$---are probed
using experimental observables that are sensitive to their
variation. In this way correlations between the density slope of the
symmetry energy $L(\rho_0)$ and a multitude of isovector observables
have been established, with neutron skins of heavy nuclei and radii
of neutron stars~\cite{Brown:2000,Horowitz:2000xj,Furnstahl:2001un}
exhibiting notably strong correlations with $L$.

In this paper we begin examining the density dependence of these
correlations, as well as the origin of the correlations by
decomposing $L$ in terms of quantities that have a more direct
physical meaning in finite nuclei and thus open up further potential
experimental probes. By employing a least-squares covariance
analysis with the correlation coefficient defined as
\begin{equation}
C_{AB} = \frac{{\rm Cov}(AB)}{{\rm Var}(A){\rm Var}(B)}  \ ,
\end{equation}
we provide for the first time a proper statistical measure of
correlations~\cite{Reinhard:2010wz,Fattoyev:2011ns,Fattoyev:2012rm}
between the density slope of the symmetry energy {\sl as a function
of baryon density} $L(\rho)$  and a selected number of isovector
observables: (a) the neutron skin thickness, (b) radii of neutron
stars and (c) the crust-core transition pressure. We will discuss
the emergence of these correlations by decomposing the $L(\rho)$ in
terms of the single-nucleon potentials in asymmetric nuclear matter
as shown in Refs.~\cite{Xu:2010fh,Xu:2010kf,Chen:2011ag}. While in
general the correlation coefficient $C_{AB}$ at a given density may
not be able to assess systematic errors reflecting limitations of
the model, the strongest correlation coefficient of almost $+1$ (or
anticorrelation of almost $-1$) at a particular density should
deliver a more universal model-independent message. For this reason,
we scan the correlation coefficients between the isovector
observables and the $L(\rho)$ over a wide range of baryon densities.

Based on the Hugenholtz-Van Hove theorem~\cite{Hugenholtz:1958zz},
it was shown that both the nuclear symmetry energy $S(\rho)$ and its
density slope $L(\rho)$ can be decomposed in terms of the
single-nucleon potentials~\cite{Chen:2011ag}. For convenience, we
will rewrite those expressions here as $S(\rho) = S_1(\rho) +
S_2(\rho)$ and $L(\rho) = L_1(\rho) + L_2(\rho) + L_3(\rho)
+L_4(\rho) + L_5(\rho) $, where
\begin{eqnarray}
&& L_1(\rho) = \frac{2 \hbar^2 k_{\rm F}^2}{6 m_0^{\ast}(\rho,
k_{\rm F})} \equiv 2 S_1(\rho) \ \\
&& L_2(\rho) = -\frac{ \hbar^2 k_{\rm F}^3}{6 m_0^{\ast 2}(\rho,
k_{\rm F})} \frac{\partial m_0^{\ast}(\rho, k)}{\partial
k}\bigg|_{k = k_{\rm F}}  \\
&& L_3(\rho) = \frac{3}{2} U_{\rm sym, 1}(\rho, k_{\rm F}) \equiv 3
S_2(\rho) \ \\
&& L_4(\rho) = \frac{\partial U_{\rm sym, 1}(\rho, k)}{\partial k}\bigg|_{k = k_{\rm F}} \cdot k_{\rm F} \ \\
&& L_5(\rho) = 3 U_{\rm sym, 2}(\rho, k_{\rm F}) \ .
\end{eqnarray}
The expressions above are valid at arbitrary baryon densities, where
$m_0^{\ast}(\rho, k)$ is the nucleon effective mass in symmetric
nuclear matter (SNM), while $U_{\rm sym, 1}$ and $U_{\rm sym, 2}$
are the first- and the second-order nuclear symmetry potentials
defined as:
\begin{eqnarray} \nonumber
U_{\rm sym, i}(\rho, k) &\equiv& \frac{1}{i\!}\frac{\partial^i
U_{\rm n}(\rho, \alpha, k)}{\partial \alpha^i} \Bigg|_{\alpha = 0} =
\ \\ &=& \frac{(-1)^i}{i\!}\frac{\partial^i U_{\rm p}(\rho, \alpha,
k)}{\partial \alpha^i} \Bigg|_{\alpha = 0} \ .
\end{eqnarray}
Here $U_{\rm n}$ and $U_{\rm p}$ are the single-neutron and
single-proton potentials respectively, which generally depend on the
baryion density $\rho$, the isospin asymmetry $\alpha$ and the
amplitude of the nucleon momentum $k$. Physically, $S_1(\rho)$ and
accordingly $L_1(\rho)$ represent the kinetic energy part of the
symmetry energy that includes the isocalar effective mass
contribution, $L_2(\rho)$ describes the momentum dependence of the
nucleon effective mass, $S_2(\rho)$ hence $L_3(\rho)$ are due to the
first-order symmetry potential contribution, $L_4(\rho)$ comes from
the momentum dependence of the first-order symmetry potential, and
$L_5(\rho)$ comes from the second-order symmetry potential. Since by
definition, $L(\rho) \equiv 3\rho\frac{\partial S(\rho)}{\partial
\rho}$, it is then obvious that there is also a required closure
relation between the density derivative of the first-order symmetry
potential $U_{{\rm sym}, 1}$ and the magnitude of $U_{{\rm sym},
2}$, in particular.

\begin{figure}[ht]
\smallskip
 \includegraphics[width=8.5cm,angle=0]{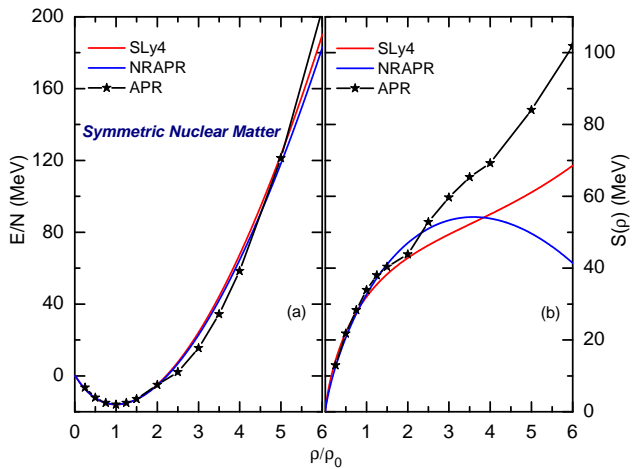}
\caption{(color online). Binding energy per nucleon in SNM (a) and
the symmetry energy (b) as a function of baryon density
$\rho/\rho_0$ for SLy4 and NRAPR Skyrme EDFs.} \label{Fig1}
\end{figure}

\begin{figure}[ht]
\smallskip
 \includegraphics[width=8.5cm,angle=0]{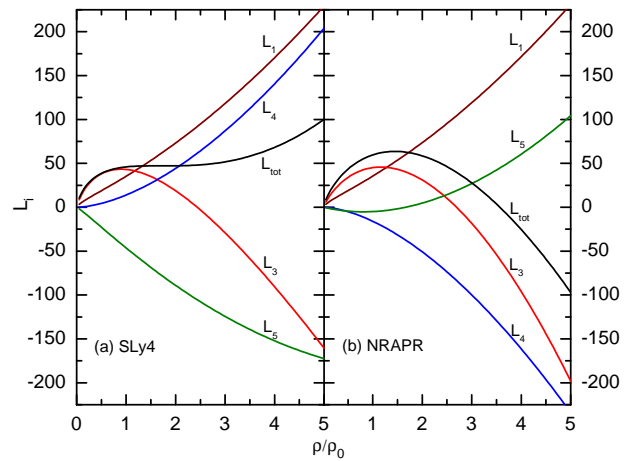}
\caption{(color online). Density dependence of the $L(\rho)$ and its
decomposition for SLy4 and NRAPR Skyrme EDFs.} \label{Fig2}
\end{figure}

The decomposition method considered above is quite general. In this
exploratory study, we report our results for two Skyrme energy
density functionals (EDFs) SLy4~\cite{Chabanat:1997qh} and
NRAPR~\cite{Steiner:2004fi} that have been widely used in the
literature, both nuclear- and astrophysical. Since our aim is to
study the isospin-dependent properties of asymmetric nuclear matter,
SLy4 and NRAPR are natural choices for the following reasons. First,
they predict an almost identical EOS of SNM (see the (a) panel of
Fig.~\ref{Fig1}). Therefore, the $L_1(\rho)$ term of the density
slope is almost indistinguishable for these interactions due to the
equivalence of the isoscalar effective mass, and the nuclear
saturation density $\rho_0$ (see Fig.~\ref{Fig2}). Notice however
that since the nucleon effective mass in Skyrme EDFs does not depend
on momentum, the $L_2(\rho)$ term becomes identically zero. Second,
both Sly4 and NRAPR reproduce the symmetry energy predicted by the
APR EOS~\cite{Akmal:1998cf} up to $\sim 1.5\rho_0$ (see the (b)
panel of Fig.~\ref{Fig1}). Whereas a similar density dependence of
$L_3(\rho)$---hence $S_2(\rho)$ or $U_{{\rm sym}, 1}$---is observed
in these interactions, both $L_4(\rho)$ and $L_5(\rho)$ terms have a
completely different density dependence (see Fig.~\ref{Fig2}).
Consequently these models have a totally different density
dependence of the symmetry energy at supra-saturation densities of
$\rho \gtrsim 1.5\rho_0$.

As a starting point, using the currently accepted uncertainty ranges
by the community we fix the isoscalar properties of bulk nuclear
matter such as the nuclear saturation density $\rho_0$, the binding
energies per nucleon in SNM at saturation $B(\rho_0)$ and at twice
saturation density $B(2\rho_0)$, the incompressibility coefficient
of nuclear matter $K_0$, the nucleon isoscalar effective mass
$m_{\rm s}^{\ast}(\rho_0)$, and the macroscopic gradient coefficient
$G_{\rm s}$~\cite{Chabanat:1997qh} at a 2\% level, while allow the
isovector effective mass $m_{\rm v}^{\ast}(\rho_0)$, and the
symmetry-gradient coefficient $G_{\rm v}$ to have a 20\% theoretical
error-bars~\cite{Chen:2010qx}. We do not {\sl a priory} assume any
error-bars on the symmetry energy parameters, hence they have not
been included in the $\chi^2$. Rather we include a conservative
range of theoretical data points for the neutron-matter energy at
densities $0.04 < \rho/\rho_0 < 0.12$ fm$^{-3}$ that were calculated
using quantum Monte Carlo calculations with chiral effective field
theory interactions~\cite{Gezerlis:2013ipa}.

\begin{figure}[ht]
\smallskip
\includegraphics[width=8.5cm]{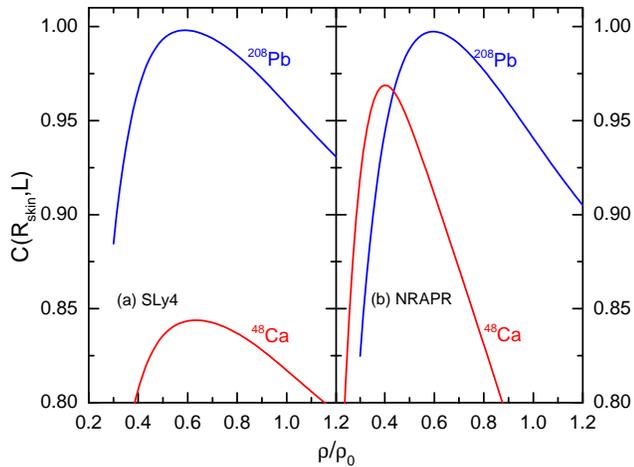}
\caption{(color online). Correlation coefficients between the
$L(\rho)$ and the neutron skin thicknesses of $^{208}$Pb and
$^{48}$Ca calculated using SLy4 (left) and NRAPR (right) Skyrme
EDFs.} \label{Fig3}
\end{figure}

\begin{figure}[ht]
\includegraphics[width=8.5cm]{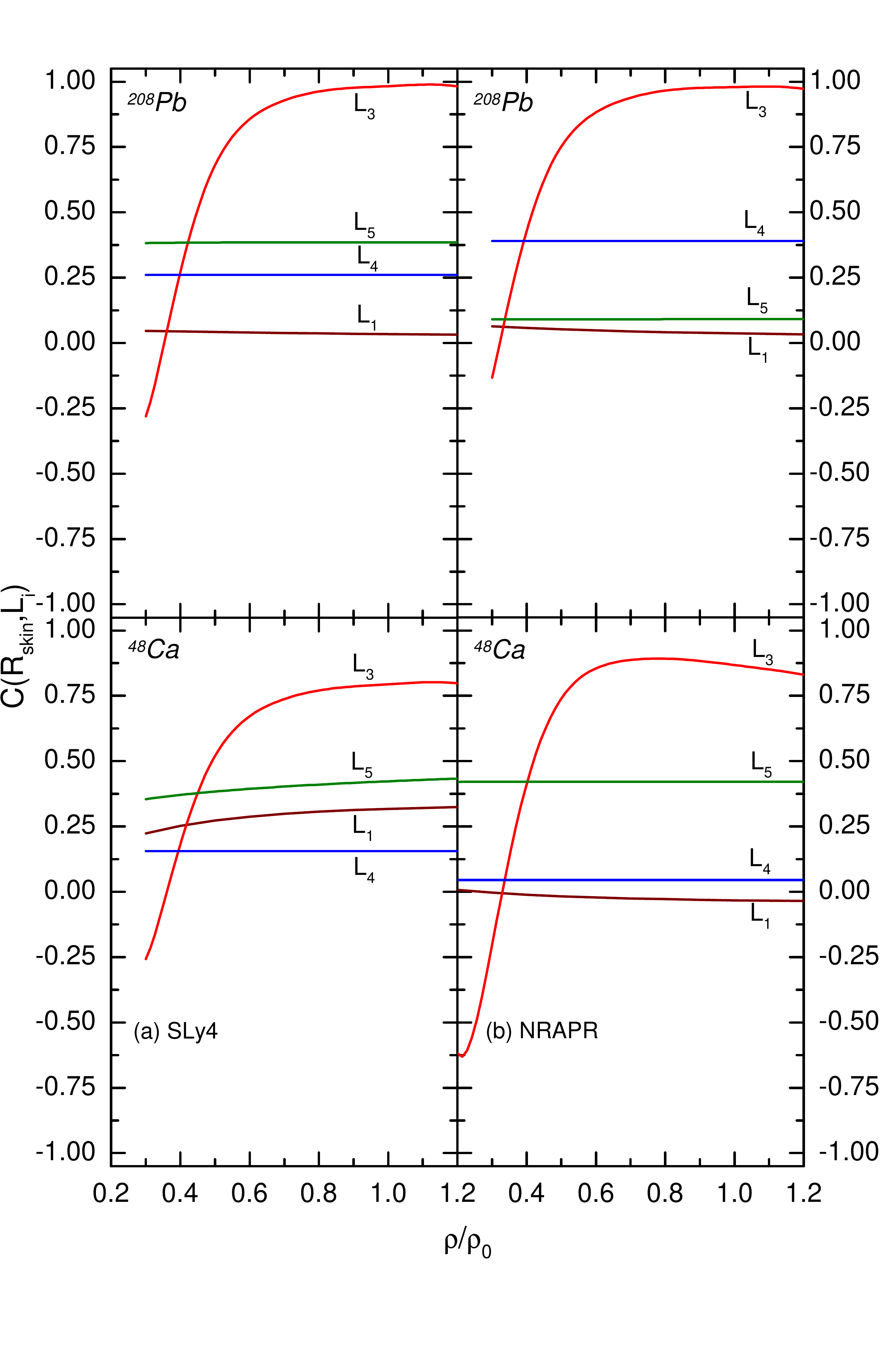}
\vskip -1 cm \caption{(color online). Correlation coefficients
between various terms in the $L(\rho)$ and the neutron skin
thicknesses of $^{208}$Pb (upper) and $^{48}$Ca (lower) calculated
using SLy4 (left) and NRAPR (right) Skyrme EDFs.} \label{Fig4}
\end{figure}

In Fig.~\ref{Fig3} we display the correlation coefficients between
the $L(\rho)$ as a function of the baryon density and the neutron
skin thicknesses of $^{208}$Pb and $^{48}$Ca. An updated measurement
of the skin thickness in $^{208}$Pb, and a proposal for measuring
the skin thickness in $^{48}$Ca has been recently approved at the
Thomas Jefferson National Accelerator
Facility~\cite{Horowitz:2013wha}. As evident from the figure,
although the strong correlation between $R_{\rm skin}(^{208}{\rm
Pb})$ and the slope of the symmetry energy at saturation $L(\rho_0)$
is present consistent with previous studies, the strongest
correlation coefficient of almost $+1$ appears only at a much lower
density of about $\rho/\rho_0 = 0.59$ for both EDFs. This means that
a measurement of the neutron skin in $^{208}{\rm Pb}$ would uniquely
determine the slope of the symmetry energy at this particular
sub-saturation density. Indeed, a recent systematic study also
showed that a strong correlation coefficient between $R_{\rm
skin}(^{208}{\rm Pb})$ and $L(\rho)$ emerges at a sub-saturation
cross density of $\rho_{\rm c} \approx 0.11$
fm$^{-3}$~\cite{Zhang:2013wna}. This result should not come as a
surprise, since only about one-third of the nucleons in $^{208}{\rm
Pb}$ occupy the saturation density area, which therefore explains
why the neutron skin thickness should constrain the $L(\rho)$ not at
$\rho_0$, but at a characteristic density in finite nuclei, which is
localized close to a mean value of the density of
nuclei~\cite{Khan:2012ps}. One should note that the neutron skin is
formed as a result of the competition between the surface tension
and the pressure of neutrons in heavy nuclei, the latter being
closely related to the $L(\rho)$. Hence the greater is the $L(\rho)$
the thicker is the skin~\cite{Brown:2000,
Horowitz:2000xj,Furnstahl:2001un}. This simple picture is of course
relevant for heavy nuclei, where the mean-field approximation is
adequate. In the case of light nuclei such as $^{48}{\rm Ca}$ the
mean-field approximation may not be sufficient, and one should
expect beyond mean-field effects to crop up. Indeed, as seen in
Fig.~\ref{Fig3} although the correlation remains as strong at a
sub-saturation density, it is not close to +1, leaving a very rich
unexplored physics behind. Moreover, the strongest correlation
coefficients occur at different densities of $\rho = 0.63 \rho_0$
for SLy4 and at $\rho = 0.39 \rho_0$ for NRAPR. Thus, simultaneous
measurement of $R_{\rm skin}$ in both $^{208}{\rm Pb}$ and
$^{48}{\rm Ca}$, are very important as they do not only help map out
the density dependence of the symmetry energy in a broader
subsaturation density region, but should also provide complementary
information on the structure of neutron-rich calcium isotope
$^{48}{\rm Ca}$~\cite{Horowitz:2013wha}. Next, in Fig.~\ref{Fig4} we
display correlation coefficients between $R_{\rm skin}$ and the
individual decomposed terms of $L(\rho)$ at various sub-saturation
densities. Except $L_3(\rho)$, or $U_{{\rm sym}, 1}$, all other
terms show almost no sign of a correlation. In fact, one should not
expect any correlation with $L_1(\rho)$ term as it is purely
isocalar in nature. The little correlation with $L_4(\rho)$ and
$L_5(\rho)$ indicate that the size of the neutron skin is not
particularly sensitive to the second-order symmetry potential and
the momentum dependence of the first order symmetry potential.

\begin{figure}[ht]
\smallskip
 \includegraphics[width=8.5cm,angle=0]{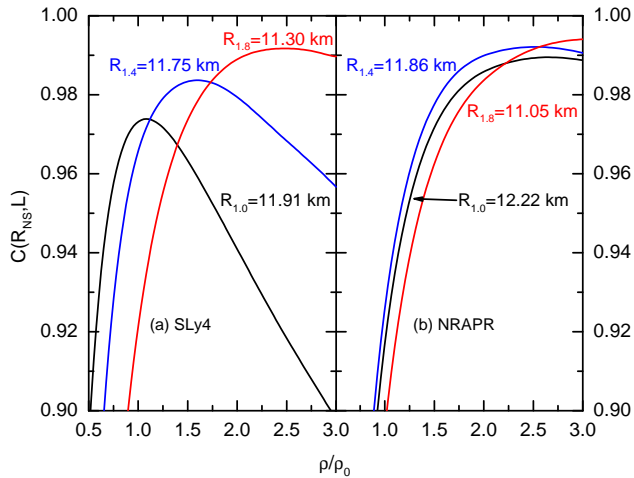}
 \caption{(color online). Correlation coefficients between $L(\rho)$
and the $1.0$-, $1.4$-, and $1.8$-solar mass neutron star radii as a
function of the baryon density calculated using SLy4 (left) and
NRAPR (right) Skyrme EDFs.} \label{Fig5}
\end{figure}

\begin{figure}[ht]
\smallskip
\includegraphics[width=8.5cm,angle=0]{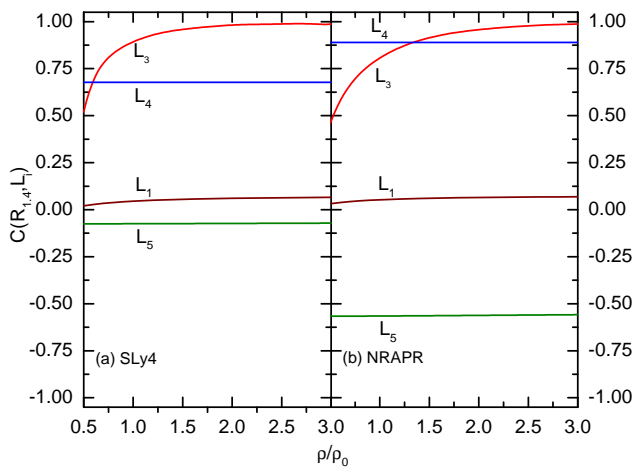}
\caption{(color online). Correlation coefficients between various
terms in the $L(\rho)$ as a function of the baryon density and a
canonical $1.4$-solar mass neutron star radii are calculated using
SLy4 (left) and NRAPR (right) Skyrme EDFs.} \label{Fig6}
\end{figure}

Since the pressure of neutron-rich matter also supports neutron
stars against gravitational collapse albeit at different densities
than found at the center of finite nuclei, one should also expect
the strong correlation to emerge between the $L(\rho)$ and the
neutron star radii. This correlation is complicated by the fact that
the radius of a neutron star samples the symmetry energy over a
range from half saturation density up to several times saturation
density; one should expect potentially significant variations in the
correlation coefficient as a function of density from EOS to EOS. We
demonstrate this by displaying the correlation coefficients between
the neutron star radii and the $L(\rho)$ as a function of density
for both SLy4 and NRAPR models in Fig.~\ref{Fig5}. In the case of
SLy4, the radius of a 1.0-solar mass neutron star shows a strong
correlation with the density slope at saturation. As the mass of the
neutron star increases, the strongest correlation shifts to the
$L(\rho)$ at higher densities, {\sl e.g.} at $1.5\rho_0$ for a
1.4-solar mass neutron star, and at $2.5\rho_0$ for a 1.8-solar mass
neutron star. Moreover, the correlation coefficient remains almost
flat for higher densities in a 1.8-solar mass neutron star. Higher
mass stars sample the internal pressure at a higher average density.
SLy4 has monotonically increasing $L(\rho)$, and thus the internal
pressure at higher densities will continue to be dominated by the
symmetry energy contribution.

NRAPR exhibits a different evolution of the correlation coefficient
with neutron star mass, with a much less pronounced increase in its
peak as we move to higher masses. The correlation for 1.0 and
1.4-solar mass stars peak at a densities of around $\sim 2.5 \rho_0$
while the correlation for a 1.8-solar mass star peaks at the
slightly higher density of $\sim 3.5 \rho_0$. $L(\rho)$ at
supra-saturation densities for NRAPR is non-monotonic, peaking at
around 1.5$\rho_0$ before falling to zero at $3.5\rho_0$---the
density range within which the radius shows its peak correlation
with $L$. Beyond the peak $L(\rho)$, the contributions to the
internal pressure from the higher-order symmetry coefficients and
from the symmetric part of the EOS will become steadily more
important, thus quenching the correlation with $L$. The two
behaviors of $L(\rho)$ exhibited by the two Skyrme models here
broadly bracket the types of behaviors seen in all Skyrme models,
and the behaviors of the correlation coefficient between radius and
$L(\rho)$ with increasing NS mass can be expected to similarly
bracket the range of possible behaviors in such models.
Nevertheless, for both models the strongest correlation coefficient
for a 1.4-solar mass star emerges with $L(\rho)$ at supra-saturation
densities, between $1.5 \rho_0$ and $2 \rho_0$. The need for the
symmetry energy in this range for the determination of neutron star
radii was also empirically observed in Ref.~\cite{Lattimer:2006xb}.
Thus measurements of the neutron skin in finite nuclei and the
radius of neutron stars probe quite different density regimes,
unsurprisingly. The latter of these regimes may be tested with
collisions of very neutron-rich heavy ions at different beam
energies~\cite{Horowitz:2014bja}.

We show the correlation coefficients between the radius of a
1.4$M_{\odot}$ star and the components of $L(\rho)$ in
Fig.~\ref{Fig6}. Similar to the neutron skin case, no correlation is
found with $L_1(\rho)$, and relatively mild correlations or
anti-correlations are found with $L_4(\rho)$ and $L_5(\rho)$. It is
again the $L_3(\rho)$ term that appears to have strongest
correlation at higher densities. Recall that $L_3(\rho) =
\frac{3}{2} U_{{\rm sym}, 1}$. This indicates that refinements in
extracting $U_{{\rm sym}, 1}$ from the measurement of the nucleon
optical model potentials, and from heavy ion collision observables,
would improve our predictions of neutron skins and neutron star
radii.

\begin{figure}[ht]
\smallskip
\includegraphics[width=8.5cm,angle=0]{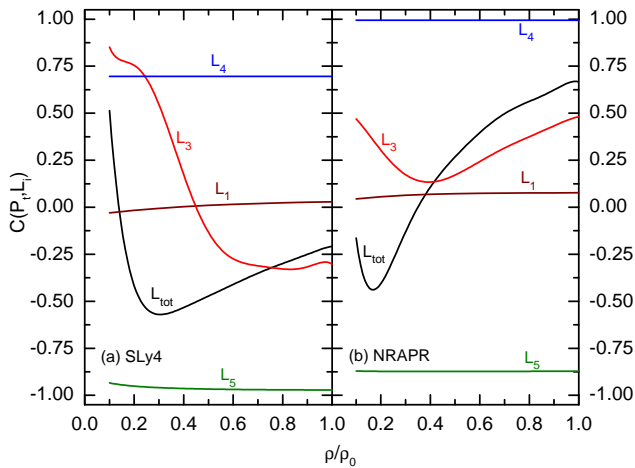}
\caption{(color online). Correlation coefficients between various
terms in the $L(\rho)$ and the crust-core transition pressure are
calculated using SLy4 (left) and NRAPR (right) models.} \label{Fig7}
\end{figure}

Finally, we discuss our results for the crust-core transition
pressure, the most important correlate with the thickness, mass, and moment
of inertia of neutron star's crust~\cite{Fattoyev:2010tb, Piekarewicz:2014lba}. 
Several methods have been used to determine the crust-core
transition properties~\cite{Horowitz:2000xj, Xu:2009vi,
Avancini:2010ch, Ducoin:2011fy}. Unlike the neutron skin thickness
and the radii of neutron stars, the crust-core transition pressure
cannot be determined entirely by the pressure of pure-neutron rich
matter itself. Indeed, in the simplest case of the thermodynamical
approach the following mechanical stability condition,
$\left(\frac{\partial P}{\partial \rho}\right)_{\mu} > 0$,
must be satisfied in order for the system to be stable against small
density fluctuations, where $\mu$ is the chemical potential. 
Since the density derivative of the pressure is quite complicated, a
complex correlation between the transition pressure and the
$L(\rho)$~\cite{Li:1997ra, Xu:2009vi, Fattoyev:2010tb,
Ducoin:2011fy} must therefore emerge. We determine the transition
pressure from a compressible liquid drop model of the crust outlined
in Ref.~\cite{Newton:2011dw}.  By plotting the correlation
coefficients between various terms in the $L(\rho)$ and the
crust-core transition pressure in Fig.~\ref{Fig7} we observe that
the values of $L_4(\rho)$ and $L_5(\rho)$---that is the momentum
derivative of $U_{{\rm sym}, 1}$ and the magnitude of $U_{{\rm sym},
2}$---are most important in the determination of the transition
pressure. The complex correlation between the crust-core transition
pressure and the $L(\rho)$ appears to originate from the complicated
behavior of the correlation with $L_3(\rho)$ coupled with the
balance between the strong correlation between the crust-core
transition pressure and $L_4(\rho)$, and the similarly strong
anti-correlation with $L_5(\rho)$, which together complicate the
emergence of correlation with the total density slope. Thus
extracting the density and momentum dependence of $U_{{\rm sym},
1}$, and the value of $U_{{\rm sym}, 2}$ are both very crucial in
determining the crust-core transition pressure, which plays critical
role in understanding many phenomena related to the neutron star
crust~\cite{Newton:2014tga}.

In summary, we have quantitatively mapped the correlations between
$L(\rho)$ and the neutron skin thickness, radii of neutron stars and
the crust-core transition pressure over a wide range of densities
for two widely used Skyrme models SLy4 and NRAPR which have similar
symmetric nuclear matter EOSs but diverging behaviors of $L(\rho)$
at supra-saturation densities. We have also calculated the
correlation coefficients between the symmetry potential component of
$L$ at saturation density and the same set of isovector observables.
We found that for the neutron skin thickness of $^{208}$Pb the
strongest correlation appears at a subsaturation density of $\rho =
0.59 \rho_0$, and that the origin of this correlation is found to be
tied to the magnitude of the first-order symmetry potential $U_{\rm
sym,
1}$. 
A similarly strong correlation exists with the radius of neutron
stars and $L(\rho)$ over a wide range of supra-saturation densities.
The behavior of the correlation is seen to depend on the behavior of
$L(\rho)$ at supra-saturation densities. If  $L(\rho)$ continues to
monotonically increase, then the peak correlation of radius with $L$
occurs at higher densities for higher mass stars. If  $L(\rho)$
increases to a maximum and starts decreasing above a certain
supra-saturation density, then the peak correlation of radius with
$L$ occurs within a similar density range independent of mass.
The radius is also found to correlate most strongly with the
magnitude of the first-order symmetry potential $U_{\rm sym, 1}$.

The crust-core transition pressure, on the other hand, is found to
be strongly correlated not with the value of $U_{{\rm sym}, 1}$ but
with its momentum derivative term and the magnitude of $U_{{\rm
sym}, 2}$. Future improvements in extracting the density and
momentum dependence of first- and second-order symmetry potentials
from optical model analysis of nucleon-nucleus scattering and at
facilities for rare isotope beams will therefore provide strong
constraints on the density dependence of the symmetry energy.

\begin{acknowledgments}
We would like to thank Prof. Lie-Wen Chen and Prof. Jorge
Piekarewicz for fruitful discussions. This work was supported in
part by the National Aeronautics and Space Administration under
Grant No. NNX11AC41G issued through the Science Mission Directorate,
and the National Science Foundation under Grant No. PHY-1068022.
\end{acknowledgments}

\bibliography{ReferencesFJF}
\vfill\eject
\end{document}